\documentclass[12pt]{article}
\usepackage{graphicx,cite}
\begin{document}

\begin{center}
{\Large\bf Chaos in de Broglie - Bohm quantum mechanics \\
and the dynamics of quantum relaxation}
\vskip 1cm
{\bf C. Efthymiopoulos, G. Contopoulos and A.C. Tzemos}\\
Research Center for Astronomy and Applied Mathematics, Academy of Athens\\
Soranou Efessiou 4, 115 27 Athens, Greece\\
emails: cefthim@academyofathens.gr, gcontop@academyofathens.gr, 
thanasistzemos@gmail.com
\end{center}
\vskip 1cm

\noindent
{\small
{\bf Abstract:}
We discuss the main mechanisms generating chaotic behavior of the quantum 
trajectories in the de Broglie - Bohm picture of quantum mechanics, in 
systems of two and three degrees of freedom. In the 2D case, chaos is
generated via multiple scatterings of the trajectories with one or more 
`nodal point - X-point complexes'. In the 3D case, these complexes form 
foliations along `nodal lines' accompanied by `X-lines'. We also identify 
cases of integrable or partially integrable quantum trajectories. The role 
of chaos is important in interpreting the dynamical origin of the `quantum 
relaxation' effect, i.e. the dynamical emergence of Born's rule for the 
quantum probabilities, which has been proposed as an extension of the 
Bohmian picture of quantum mechanics. In particular, the local scaling laws 
characterizing the chaotic scattering phenomena near X-points, or X-lines, 
are related to the global rate at which the quantum relaxation is observed 
to proceed. Also, the degree of chaos determines the rate at which 
nearly-coherent initial wavepacket states lose their spatial coherence 
in the course of time.}

\section{Introduction}

In recent years, the de Broglie - Bohm picture of quantum mechanics 
\cite{debro1928}\cite{bohm1952} attracted the interest of researchers in various 
subfields of quantum physics. Besides an ongoing debate on its ontological 
significance \cite{bohmhil1993}\cite{hol1993}\cite{duretal996}\cite{cush2000}\cite{durteu2009}\cite{val2014}, the Bohmian approach has led to the development 
of new calculational tools and trajectory-based methods \cite{lopwya1999}
\cite{wya2005}\cite{plamom2012}\cite{bensetal2014}\cite{sanmir2014} which 
allow to study quantum phenomena from new and challenging perspectives.  

In its standard formulation, the de Broglie - Bohm picture yields predictions 
equivalent to those of the Schr\"{o}dinger and Heisenberg picture of quantum 
mechanics. On the other hand, Valentini \cite{val1991}\cite{peaval2006} proposed 
a non-standard extension of the de Broglie - Bohm picture, acccording to which, 
the ordinary quantum mechanics represents only a limiting case of physical systems, 
i.e., systems which are close to a state of quantum equilibrium. In such a state, 
Born's rule $p=|\psi|^2$ is satisfied, where $p$ is the probability of an outcome 
of a particular measurement, and $\psi$ is the wavefunction. However, Born's rule 
is regarded not as an axiom or condition implied to the whole universe, but rather 
as an emergent property arising {\it dynamically} from the collective effect of 
ensembles of Bohmian trajectories. Via the `sub-quantum H-theorem' \cite{val1991}, 
it is asserted that, under particular assumptions for the initial wavefunction, 
the Bohmian trajectories undergo {\it quantum relaxation}, i.e., a gradual 
approach of the probability $p$ towards the limit $p\rightarrow|\psi|^2$, 
which occurs as $t\rightarrow\infty$, even when the initial probability $p_0$ 
is different from $|\psi_0|^2$. 

A key feature of quantum relaxation is that, in principle, it can occur even 
in small systems of few degrees of freedom which are considered as practically 
isolated from their environment. This comes in contrast to the so-called typicality 
condition \cite{duretal1992} which can justify Born's rule in small systems via 
the notion of `conditional wavefunction', which, however, applies only when a 
system is considered coupled to the environment \cite{duretal1992}. 

On the other hand, numerical studies in particular systems \cite{valwes2005}
\cite{eftcon2006}\cite{colstru2010}\cite{conetal2012}\cite{col2012}
\cite{eitetal2014}\cite{kanval2016} have given both examples and counter-examples 
of the occurence of quantum relaxation. A common feature in all such studies is 
the identification of the key role played in the quantum relaxation process by 
the underlying degree of complexity of the individual Bohmian trajectories. 
In fact, systems with regular trajectories undergo limited quantum relaxation, 
thus their quantum probabilities exhibit fluctuations with respect to 
Born's rule. According to Valentini (see \cite{kanval2016} and references 
there in), such fluctuations are indeed expected in some very special systems. 
Of particular interest are the fluctuations of quantum particles and fields in 
the early universe, which would leave observable signatures in high precision 
cosmological or astrophysical data \cite{val2007}\cite{val2009}\cite{val2010}\cite{colval2013}\cite{colval2015}\cite{undval2015}\cite{undval2016}. 

In the present review we provide a summary of results on the dynamics 
of individual Bohmian trajectories and its relation to the dynamical origin 
of quantum probabilities via the quantum relaxation mechanism. In a series 
of works \cite{eftcon2006}\cite{eftetal2007}\cite{coneft2008}\cite{conetal2008}
\cite{eftetal2009}\cite{conetal2012}\cite{tzeetal2016} we emphasized the 
importance, in this process, of the phenomenon of {\it chaos} (see also 
\cite{ben2010}). Chaos manifests itself via the sensitive dependence 
of the quantum trajectories on the initial conditions, quantified by 
positive values of the trajectories' Lyapunov characteristic exponents. 
The existence of chaotic Bohmian trajectories is well established in the 
literature \cite{duretal1992}\cite{faisch1995}\cite{parval1995}\cite{depol1996}
\cite{dewmal1996}\cite{iacpet1996}\cite{fri1997}\cite{konmak1998}\cite{wuspru1999}
\cite{maketal2000}\cite{cush2000}\cite{desalflo2003}\cite{falfon2003}
\cite{wispuj2005}\cite{wisetal2007}\cite{schfor2008}\cite{cesstru2016}. 
In \cite{eftcon2006}\cite{conetal2012}\ we argued that the chaotic behavior 
of the Bohmian trajectories is a necessary condition for a system to exhibit 
quantum relaxation. In particular, chaos generates a so-called `mixing' property 
of the trajectories \cite{eftcon2006}, which is a dynamical effect analogous to 
the effect of hypothetical `sub-quantum' forces conjectured in an older theory 
by Bohm and Vigier \cite{bohmvig1954}. It should be stressed, however, that 
contrary to the Bohm - Vigier theory, chaos is a dynamical property of the 
Bohmian trajectories under the ordinary quantum evolution. This means that the 
appearance of chaos leads to quantum relaxation without requiring any new 
asumption beyond the ordinary assumptions of standard quantum mechanics. 

In our works we studied how generic the appearance of chaos is in the Bohmian 
trajectories in systems of two and three degrees of freedom, as well as the 
dynamical mechanisms of chaos. Wisniacki and Pujals \cite{wispuj2005}
\cite{wisetal2007} first made the observation that chaos is generated in systems 
possessing {\it moving quantum vortices}. These are particular structures of the 
quantum flow formed around moving nodal points, i.e., points in configuration 
space where the wavefunction becomes equal to zero. 
In \cite{eftetal2007}\cite{eftetal2009} we showed that the form of the 
quantum-mechanical equations of motion imposes a general structure of the quantum 
flow around a nodal point, which we call `nodal point - X-point complex' (where 
the `X-point' is an unstable stationary point of the instantaneous quantum flow 
close to the nodal point, see section 2 below). This structure is responsible for 
the chaotic scattering of the Bohmian trajectories, as analyzed theoretically in 
\cite{eftetal2009}. This analysis allows also to quantify the local value of the 
Lyapunov number of a trajectory at each scattering event. The results in the above 
works, applying to 2D systems, were recently generalized in the case of 3D systems 
\cite{tzeetal2016}. In the 3D case, the foliation of nodal point - X-point 
complexes forms a 3D cylindrical structure, which contains a `nodal line' 
(discussed already in \cite{wuspru1999}\cite{falfon2003}), but also an `X-line'. 
The addition of the X-line to the 3D cylindrical structure of the quantum flow 
is important, since we find that most trajectories avoid the nodal line, while 
they have repeated close encounters with the X-line. Chaos is generated, precisely, 
at these encounters. This mechanism can be shown to be generic, i.e., independent 
of the particular wavefunction considered. Moreover, special cases, where 
the motion of the nodal points may not be necessary for the appearance of chaos, 
are studied in \cite{cesstru2016}. 

On the other hand, in our studies we found also that there are many cases of 
ordered Bohmian trajectories, which can be represented by exact or approximate 
integrals of motion \cite{coneft2008}\cite{conetal2012}\cite{tzeetal2016}
\cite{conetal2016s} and exhibit no chaos. In 2D systems, we found cases 
where approximate integrals of motion can be constructed in the form of power 
series \cite{coneft2008}\cite{conetal2012}, while in 3D systems we found several 
cases of systems possessing exact integrals of motion \cite{tzeetal2016}
\cite{conetal2016s}. The existence of regular trajectories obstructs quantum 
relaxation, putting restrictions to the overall mixing of the trajectories. 
In fact, the co-existence of order and chaos is a general feature found in 
most examples studied so far in the literature. It is unknown, however, 
to what extent chaos prevails in more general systems with an increasingly 
complicated quantum state, for which the rate of approach to quantum equilibrium 
has so far been studied only by numerical means \cite{towetal2012}\cite{eitetal2014}. 

As a final note, besides its importance in understanding the role of chaos 
in quantum relaxation, the study of the dynamics in quantum systems close to  
singularities forming quantum vortices presents a wide spectrum of applications. 
The quantization of the action integral around quantum vortices was studied 
already by Dirac \cite{dir1931}, see \cite{hiretal1974a}\cite{hiretal1974b}. 
Some modern applications of these topics include: tunneling through potential 
barriers, \cite{hiretal1974b}\cite{skoetal1989}\cite{lopwya1999}\cite{babetal2003}, 
ballistic electron transport \cite{beeetal1991}\cite{wuspru1994}\cite{beretal2001}, 
superfluidity \cite{fey1955}, Bose-Einstein condensates \cite{dalstr1996}
\cite{rok1997}\cite{svifet1998}\cite{duazha1999}\cite{garper1999}, optical 
lattices \cite{vigetal2007}, atom-surface scattering \cite{sanzetal2004}, 
Josephson junctions \cite{bruetal1999}, decoherence \cite{nawya2002} etc. 
Finally, the study of chaos allows to quantify the precision requirements for 
so-called `hydrodynamical' methods of solution of Schr\"{o}dinger's equation 
(see e.g. \cite{wya2005}), and its recent generalizations in the study of the 
quantum N-body problem (see \cite{orietal2016} and references there in). 

In conclusion, the study of order and chaos in the quantum trajectories is a 
fruitful area of current research, with applications in a variety of theoretical 
and practical problems related to the de Broglie - Bohm picture of quantum 
mechanics. 

The paper is structured as follows: in section 2 we describe the nodal point 
X-point complexes in 2D and 3D systems, and present the mechanism of generation 
of chaos by such complexes. Section 3 refers to ordered Bohmian trajectories, 
while in section 4 we discuss the main effect on quantum relaxation by the 
interplay between ordered and chaotic trajectories. Finally, section 5 
contains the main conclusions of our study. 

\section{Nodal point - X-point complexes}

\subsection{2D quantum flow}
\begin{figure}
\centering
\includegraphics[scale=0.35]{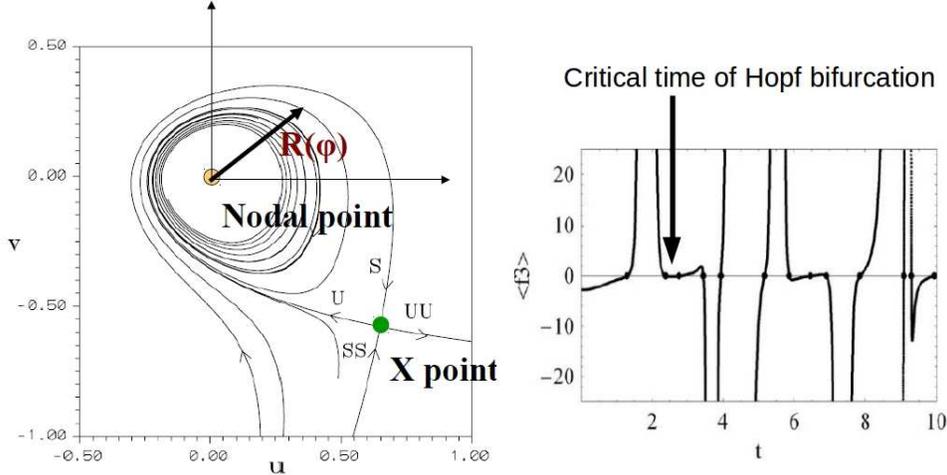}
\caption{Left panel: Basic structure of a `nodal point - X-point complex'. 
The nodal point and X-point are shown along with the asymptotic manifolds 
(unstable U,UU, stable S,SS) of the X-point. The quantum flow around 
the nodal point forms spirals. Right panel: time evolution of the quantity 
$\langle f_3\rangle$ in a specific example (see text). Changes in the sign 
of $\langle f_3\rangle$ induce a `Hopf bifurcation', which changes the 
character of the nodal point from attractor to repellor or vice versa.}
\label{fig:nodxp}
\end{figure}

Figure \ref{fig:nodxp} summarizes the basic structure of the `nodal point - X-point 
complex'. We refer to 2D systems in the co-ordinate plane $(x,y)$ with wavefunction 
$\psi(x,y,t)$. A point $(x_0,y_0)$ at the time $t$ for which $\psi(x_0,y_0,t)=0$ 
is called a nodal point of the wavefunction. We assume that $(x_0,y_0)$ is  
a simple root of the equations $Re(\psi)=Im(\psi)=0$, whose position changes 
in time $(x_0,y_0)\equiv\bigg(x_0(t),y_0(t)\bigg)$. We then set 
$\bigg(x_0(t),y_0(t)\bigg)$ at the center of a moving frame of reference, 
with instantaneous velocity $\vec{V}(t)\equiv(V_x,V_y)=(\dot{x}_0,\dot{y}_0)$, 
and we find general expressions for the structure of the quantum flow in this 
moving frame of reference. By expanding the wavefunction around the nodal 
point as
\begin{eqnarray}\label{psiexp}
\psi&=&\bigg(a_{10}(t)+ib_{10}(t)\bigg)u+\bigg(a_{01}(t)+ib_{01}(t)\bigg)v
+{1\over
2}\bigg(a_{20}(t)+ib_{20}(t)\bigg)u^2\nonumber\\
&+&{1\over 2}\bigg(a_{02}(t)+ib_{02}(t)\bigg)v^2
+\bigg(a_{11}(t)+ib_{11}(t)\bigg)u v+\ldots
\end{eqnarray}
with $u=x-x_0$, $v=y-y_0$ and real constants $a_{ij}$, $b_{ij}$, the following 
properties are seen to hold \cite{eftetal2009}: i) not all the coefficients 
$a_{10}$, $a_{01}$,$b_{10}$, $b_{01}$ can vanish at $t=t_0$, ii) the continuity 
equation $\partial\rho/\partial t + \nabla\cdot j=0$ with $\rho=|\psi|^2$, 
$j=\hbar [Re(\psi)\nabla(Im(\psi))-Im(\psi)\nabla(Re(\psi))]/(2mi)$ leads, 
to first order in $(u,v)$, to the equations
\begin{equation}\label{curfree}
a_{02}=-a_{20},~~b_{02}=-b_{20}~~.
\end{equation}
Setting Planck's constant $\hbar=1$ and the particle mass $m=1$, 
the Bohmian equations  
\begin{equation}\label{eqmoxy}
(\dot{x},\dot{y})=Im\bigg({\nabla_{x,y}\psi\over\psi}\bigg)
\end{equation}
take the following form in the moving frame of reference:
\begin{equation}\label{eqmouv}
(\dot{u},\dot{v})=Im\bigg({\nabla_{u,v}\psi\over\psi}\bigg)
-(V_x,V_y)~~.
\end{equation}
Using the above equations, we find after some algebra the equations of 
motion in the moving frame up to second degree in $u,v$:
\begin{eqnarray}\label{eqmoexp}
{du\over dt}&=&{1\over G}\times\bigg[ (a_{01}b_{10}-a_{10}b_{01})v
+ {1\over 2}(a_{02}b_{10}-a_{10}b_{02})u^2 + ({1\over
2}a_{02}b_{10}-{1\over 2}a_{10}b_{02}-a_{11}b_{01})v^2 \nonumber\\
& &+ (a_{02}b_{01}-a_{01}b_{02})u~v+\ldots\bigg]-V_x \nonumber\\
{dv\over dt}&=&{1\over G}\times\bigg[ (a_{10}b_{01}-a_{01}b_{10})u
+ {1\over 2}(a_{01}b_{02}-a_{02}b_{01})v^2 + ({1\over
2}a_{01}b_{02}-{1\over 2}a_{02}b_{01}-a_{11}b_{10})u^2\nonumber \\
& &+ (a_{10}b_{02}-a_{02}b_{10})u~v+\ldots\bigg]-V_y
\end{eqnarray}
with
\begin{equation}\label{gi}
G=(a_{10}^2+b_{10}^2)u^2+(a_{01}^2+b_{01}^2)v^2+2(a_{01}a_{10}
+b_{01}b_{10})u~v+\ldots
\end{equation}
Under certain conditions detailed in \cite{eftetal2009}, the adiabatic 
approximation holds, according to which the Bohmian trajectories locally 
follow the quantum flow obtained by `freezing' the time in the right hand 
side of Eqs.(\ref{eqmoexp}). 

\begin{figure}
\centering
\includegraphics[scale=0.4]{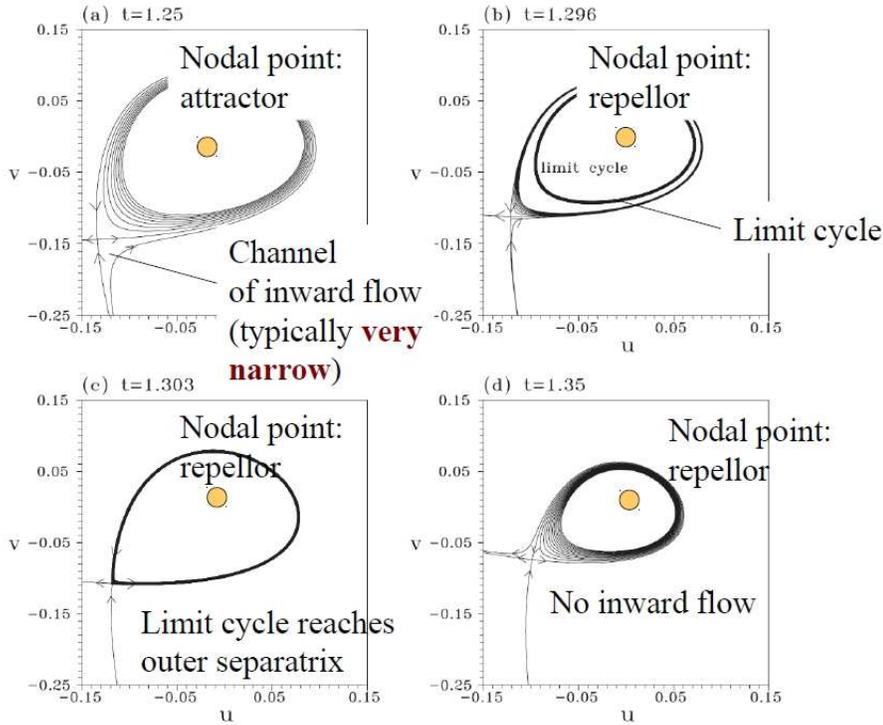}
\caption{Example of a Hopf bifurcation taking place near the nodal point 
as the time $t$ increases, referring to the data of figure 4 of \cite{eftetal2009} 
(see also \cite{eftetal2007}). (a) At $t=1.25$ the nodal point is attractor. 
One branch of the unstable manifold of the X-point forms a spiral which terminates 
asymptotically at the nodal point. (b) At $t=1.296$ a Hopf bifurcation has taken 
place. The nodal point is now a repellor, and all spirals (both from the nodal 
and the X-point) approach asymptotically a limit cycle surrounding the nodal 
point, which is an attractor.(c) At $t=1.303$ the limit cycle collides with the 
separatrix of the X-point. (d) After this collision, the limit cycle disappears 
and a transition takes place in the geometry of the asymptotic manifolds of the 
X-point. In particular, there are now no flow lines leading to the interior of 
the nodal point - X-point complex.}
\label{fig:hopf}
\end{figure}

Figure \ref{fig:nodxp} depicts the structure of the quantum flow around 
the nodal point, which has the following features:

i) {\it Nodal point.} 
Using polar coordinates $u=R\cos\phi$, $v=R\sin\phi$, Eqs.(\ref{eqmoexp}) 
are transformed to
\begin{equation}\label{eqmo2}
{dR\over dt}={c_2R^2+c_3R^3+c_4R^4+...\over G},~~~ {d\phi\over
dt}={d_0+d_1R+d_2R^2+...\over G}
\end{equation}
with coefficients $c_j$ and $d_j$ depending on the coefficients $a_{ij}$, 
$b_{ij}$, the velocities $(V_x,V_y)$, as well as powers of the trigonometric 
functions $\sin\phi,\cos\phi$. Close to the nodal point, which corresponds 
to $R=0$, we define an average value $\bar{R}$ of the radius $R$ over one 
period of revolution around the nodal point. This can now be expressed as 
a function of the azimuth $\phi$ (see Fig.\ref{fig:nodxp}) via the equation
\begin{equation}\label{drdfav}
{d\bar{R}\over d\phi}=<f_3>\bar{R}^3+...
\end{equation}
with a coefficient $<f_3>$ given by
$$
\langle f_3\rangle (a_{ij},b_{ij},V_x,V_y)={1\over 2\pi}
\int_{0}^{2\pi}\bigg({c_3\over d_0}-{c_2d_1\over d_0^2}\bigg)d\phi
$$
with $i+j=0,1,2$. The final expression for $\langle f_3\rangle$ as a function of 
the coefficients $a_{ij}$, $b_{ij}$ is given in Appendix I of \cite{eftetal2009}. 
It turns out that, depending on the sign of $\langle f_3\rangle$, the nodal point 
becomes either attractor or repellor of the quantum flow. Furthermore, as illustrated 
in the right panel of Fig.\ref{fig:nodxp}, the sign of $\langle f_3\rangle$ varies 
in time. At each change of sign, the nodal point undergoes a {\it Hopf bifurcation}, 
altering its character from attractor to repellor and vice versa. According to the 
general theory of dynamical systems, these bifurcations are accompanied by the 
generation of {\it limit cycles}, i.e. closed curves which surround the nodal 
point and act themselves as attractors or repellors of the flow. Detailed numerical 
examples of this behavior were reported in \cite{eftetal2007}. A particular example 
is reproduced here in Fig.\ref{fig:hopf}. 

ii) {\it X-point.} 
A second critical point ($u_X,v_X$) of the instantaneous flow is found 
by setting $du/dt=dv/dt=0$ at $(u,v)=(u_X,v_X)$, yielding
\begin{equation}\label{dudv}
{V_x\over V_y}=\frac{A v_X + B_1 u_X^2 + C_1 v_X^2 + D_1
u_X~v_X+\ldots} {-A u_X+ B_2 u_X^2 + C_2 v_X^2 + D_2
u_X~v_X+\ldots}
\end{equation}
with coefficients $A,B_i,C_i,D_i$ obtained from Eqs.(\ref{eqmoexp}). 
We then find that the Jacobian matrix of the linearized equations of 
motion around $(u_X,v_X)$ yield always a pair of real eigenvalues
$\lambda_1,\lambda_2$, with $\lambda_1\lambda_2<0$. In \cite{eftetal2009} 
a power-law scaling $\lambda\sim R_X^{-p}$ with $p\simeq 1.5$ was 
found for the positive eigenvalue, where $R_X = (u_X^2+v_X^2)^{1/2}$. 
As shown in Fig.\ref{fig:nodxp}, from the point $(u_X,v_X)$, called 
the `X-point', emanate two pairs of unstable (U,UU) and stable (S,SS) 
manifolds, which yield (in the adiabatic approximation) the curves 
of initial conditions of trajectories approaching asymptotically to 
the X-point in the backward and forward sense of time respectively. 

In Fig.\ref{fig:nodxp} we see an example of how the flow lines, and in 
particular the invariant manifolds emanating from the instantaneous X-point 
extend in configuration space, thus forming a complete `nodal point - X-point 
complex'. The key remark is that one of the four manifold curves (U,UU,S or 
SS) of the X-point necessarily continues as a spiral, which terminates approaching 
asymptotically either the nodal point, or the limit cycle surrounding the nodal 
point (whenever this cycle exists). Furthermore, as shown in Fig.\ref{fig:hopf}, 
the geometry of the flow lines changes in time, undergoing transitions at 
consecutive Hopf bifurcations taking place every time when the quantity 
$\langle f_3\rangle$ changes sign from positive to negative and vice versa. 
Transitions in the geometry of the flow lines also take place when the limit 
cycles, which after their generation at the nodal point expand outwards, 
arrive to a collision with the manifolds of the X-point. 

The most important remark is that, very close to the nodal points, 
the Bohmian trajectories, which follow the same structures, are 
necessarily ordered . The fact that the quantum flow very close to 
the nodal points should be ordered was noted already by Bohm (see 
\cite{falfon2003}). On the other hand, the hyperbolic character 
of the X-points implies that, in close encounters with the X-points, 
the Bohmian trajectories are chaotically scattered. This prediction 
is verified numerically by plotting the time evolution of the 
`stretching numbers' \cite{vogcon1994}, i.e., the local Lyapunov numbers of 
the trajectories as a function of the distance from an X-point. In such 
numerical experiments we observe consistently that positive local 
Lyapunov numbers develop only at the close encounters with X-points 
\cite{eftetal2007}. In \cite{eftetal2009} a theoretical estimate 
was derived of the form
\begin{equation}\label{xiovx0}
{\xi\over\xi_0}\sim {1\over V_0\delta v_1}
\end{equation}
where $\xi_0$ and $\xi$ denote the lengths of the deviation vectors
of a trajectory before and after the scattering by an X-point, $V_0$ 
denotes the velocity of the nodal point and $\delta v_1$ the `impact 
parameter', i.e., the initial distance of the trajectory from the
X-point's stable manifold. The estimate (\ref{xiovx0}) allows to 
quantify, in turn, the local value of the stretching number at every 
close encounter, which by definition is given as $\sim\ln(\xi/\xi_0)$.

\subsection{3D Quantum flow}
As demonstrated in \cite{tzeetal2016}, the concept of `nodal point - X-point' 
complex can be generalized in 3D quantum systems. An early relevant analysis 
in the 3D case was provided by Falsaperla and Fonte \cite{falfon2003}, 
based on a local expansion of the 3D Bohmian equations of motion around 
a nodal point. We note here that, in the 3D case, the nodal point equations 
$Re[\psi(x,y,z,t)]=0$, $Im[\psi(x,y,z,t)]=0$ admit, in general, as solutions,  
families of curves rather than isolated points. These curves were called 
{\it nodal lines} in \cite{falfon2003}. In fact, a whole nodal 
curve moves, in general, in the 3D configuration space as a function of 
the time $t$. 

Every nodal line contains infinitely many nodal points, and a local analysis 
has now to be made around each of these nodal points. In \cite{falfon2003} it 
was demonstrated that the streamlines of the quantum flow around one nodal point 
lie locally in a plane orthogonal to the local direction of the nodal line. 
Furthermore, the local form of the streamlines in this plane around the 
nodal point is spiral. 

\begin{figure}
\centering
\includegraphics[scale=0.5]{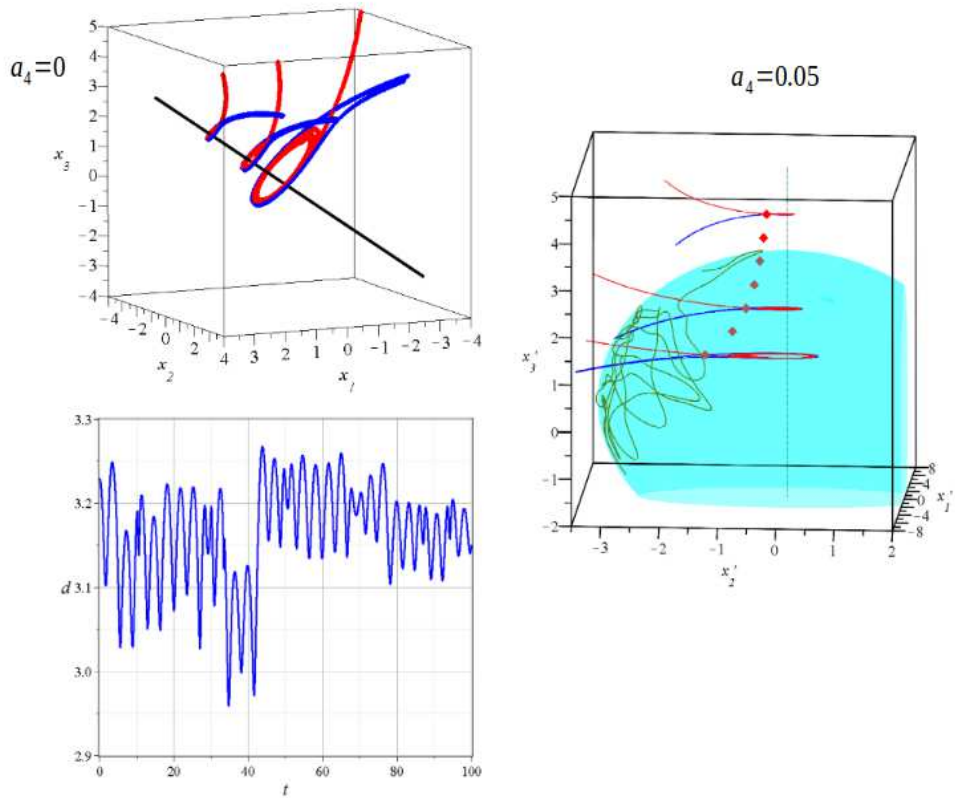}
\caption{Top left: example of the 3D structure of nodal points and X-points. 
The straight line represents the nodal line for the quantum system dealt 
with in \cite{tzeetal2016}. A foliation of planar nodal point - X-point complexes 
exists, three of which are shown in the figure. Right panel: foliation of the nodal 
point - X-point complexes along with a chaotic Bohmian trajectory undergoing 
diffusion in a direction nearly parallel to the nodal line (see \cite{tzeetal2016}). 
The red squares are points along the X-line. The time evolution of the excursions 
along the nodal line are shown in the bottom left panel.}
\label{fig:xline3d}
\end{figure}
A detailed analysis of the 3D structure of the quantum flow for generic 
wavefunctions expanded around nodal points will be presented elsewhere 
(Contopoulos et al., in preparation). The main points of this analysis were 
given in \cite{tzeetal2016} via a specific example. In summary, we found the 
following: 

Similarly to the 2D case, in each of the orthogonal planes defined 
in \cite{falfon2003}, the equations of motion can be written 
locally in co-ordinates centered around the nodal point, with directions 
parallel (coordinates $(u,v)$) or normal (coordinate $w$) to the plane. 
For the normal coordinate we find $\dot{w}\simeq 0$, while in the plane 
$(u,v)$ we obtain a set of equations of the form
\begin{equation}\label{eqmouv3d}
\dot{u} = F_1(u,v,t),~~~\dot{v} = F_2(u,v,t)
\end{equation}
for functions $F_1$ and $F_2$ having the same structure as in the r.h.s. 
of Eqs.(\ref{eqmoexp}). This implies that all results regarding the 2D nodal 
point - X-point complexes are transferable to the 3D case as well, whereas 
the entire configuration space in the 2D case corresponds now to each of the 
orthogonal planes intersecting a nodal line at each one of its nodal points.  

Figure \ref{fig:xline3d} ilustrates the cylindrical structure formed by 
taking the foliation of planes orthogonal to a nodal line, and plotting the 
nodal point - X-point complex in each of these planes. Connecting the X-points 
along such a foliation forms an `X-line' which accompanies the nodal line in 
a direction nearly parallel to it. As shown in the right panel of 
Fig.\ref{fig:xline3d}, the Bohmian trajectories are scattered when 
approaching close to the X-line. 

The main new feature with respect to the 2D case regards the speed of chaotic 
diffusion in the direction {\it along} the X-line. In several cases, we find 
that the excursions in this direction vary with time proportionally to a small 
quantity ($a_4$ in Fig.3, see \cite{tzeetal2016}) expressing the deviation of 
a system from {\it partial integrability} \cite{conetal2016s}. Several cases 
of partially integrable systems are examined in \cite{conetal2016s}. In such 
cases, quantum relaxation is obstructed by the existence of integrals of motion 
which prevent the Bohmian trajectories from diffusing in directions nearly parallel 
to those of the nodal lines. In general, quantum relaxation is obstructed whenever 
there is some form of local or global integral of motion of the quantum 
trajectories. To these integrable or partially integrable cases we now turn 
our attention. 

\section{Ordered Bohmian trajectories}

\subsection{2D case}
In 2D systems, the Bohmian trajectories appear as ordered when they never 
approach a nodal point -X point complex. Examples of ordered trajectories 
were explored in \cite{eftcon2006}\cite{eftetal2007}\cite{conetal2008}
\cite{conetal2012}. In the 2D harmonic oscillator model
\begin{equation}\label{ham2dharm}
H={1\over 2}(p_x^2+p_y^2) + {1\over 2}(x^2 + (c y)^2)
\end{equation}
we examined the Bohmian trajectories for the wavefunction 
\begin{equation}\label{eigenharm}
\psi(x,y,t) = e^{-{x^2+cy^2\over 2}-i{(1+c)t\over 2}}\big(
1+axe^{-it}+bc^{1/2}xye^{-i(1+c)t}\big)~~
\end{equation}
with real amplitudes $a,b$ and incommensurable frequencies
$\omega_1=1$, $\omega_2=c$. The equations of motion:
\begin{eqnarray}\label{eqmoha}
{dx\over dt}&=&-{a \sin t + bc^{1/2}y \sin(1+c)t \over G}\\
{dy\over dt}&=&-{bc^{1/2}x
\left(ax\sin ct +\sin(1+c)t\right) \over G}\nonumber
\end{eqnarray}
with
$$
G=1+a^2x^2+b^2cx^2y^2+2ax\cos t+ 2bc^{1/2}xy\cos(1+c)t
+ 2abc^{1/2}x^2y\cos ct
$$
admit formal series solutions applicable whenever the trajectories 
are far from the unique moving nodal point of the system, which has 
co-ordinates
\begin{equation}\label{nodal}
x_N=-{\sin(1+c)t\over a\sin ct},
~~~y_N=-{a\sin t\over bc^{1/2}\sin(1+c)t}~~.
\end{equation}
\begin{figure}
\centering
\includegraphics[scale=0.65]{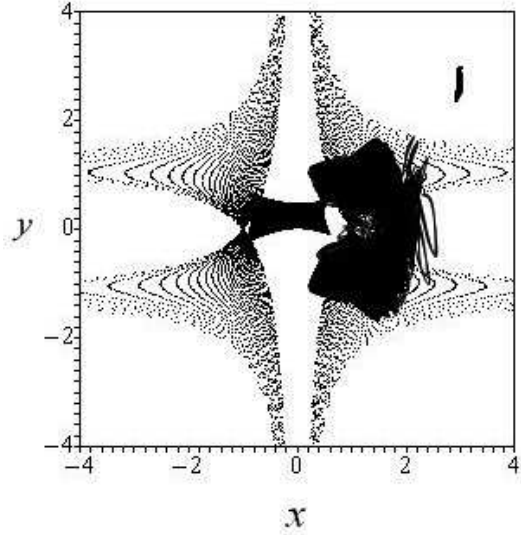}
\caption{Three Bohmian trajectories in the model (\ref{eigenharm}) for the 
same numerical data as in figure 6 of \cite{conetal2012}. The inner (central) 
and outer (upper right) trajectories are regular, and they can be described by 
formal series. The middle trajectory to the right is chaotic, and it has a 
large overlap with the domain crossed by the successive positions of the unique 
nodal point of the system (gray, corresponding to Eq.(\ref{nodal}) at different 
times $t$).}
\label{fig:hainout}
\end{figure}
In fact, as shown in Fig.\ref{fig:hainout}, it is possible to have ordered 
trajectories which partly overlap in space with the domain covered by the 
trajectory of the nodal point. The condition for a trajectory to be ordered 
is that its instantaneous position should be always far from the position of 
the nodal point. In \cite{conetal2012} this condition was investigated analytically 
for trajectories as in Fig.\ref{fig:hainout}, using formal series expansions. 
The latter are given in powers of the amplitudes $a,b$ as 
\begin{equation}\label{solx1}
x=x_0+a(\cos t-1)+{bc^{1/2}y_0\over 1+c}\bigg(\cos(1+c)t-1\bigg)+\ldots
\end{equation}
\begin{equation}\label{soly1}
y=y_0+{bc^{1/2}x_0\over 1+c}\bigg(\cos(1+c)t-1\bigg)+\ldots
\end{equation}
The series expansions are found numerically to be accurate even beyond 
the domain (in $a,b$) where the convergence of the series can be 
established by rigorous means. 

In the same model, using series expansions it is possible to demonstrate 
that ordered trajectories exist also far from the center. Considering initial 
conditions $x_0,y_0$ close to the diagonal, with $x_0,y_0$ large, we produce 
expansions of the trajectories analytical solutions in terms of the small 
quantities $1/x_0$, $1/y_0$. Setting 
\begin{equation}\label{xycap}
X(t)={1\over x_0}x(t),~~~Y(t)={1\over y_0}y(t)
\end{equation}
with $X(0)=Y(0)=1$, we then find 
\begin{equation}\label{xyexp}
X(t)=1+X_1(t)+X_2(t)+...,~~~Y(t)=1+Y_1(t)+Y_2(t)+...
\end{equation}
with $X_1=Y_1=X_2=Y_2=0$, and 
\begin{equation}\label{xysol3}
X_3(t)=0,~~~~~~~~~Y_3(t)={a(\cos ct - 1)\over bc^{3/2}y_0^3}~~,
\end{equation}
\begin{equation}\label{xysol4}
X_4={\cos(1+c)t-1\over bc^{1/2}(1+c)x_0^3y_0},~~~~~~~~~
Y_4={-a^2(\cos 2ct-1)\over 2b^2c^2y_0^4}+{\cos (1+c)t-1\over 
bc^{1/2}(1+c)x_0y_0^3}~~.
\end{equation}
Higher order terms were found by a computer-algebraic program, and numerical 
convergence tests of these series were discussed in \cite{coneft2008}
\cite{conetal2012}. 


\subsection{3D case: partial integrability}

Extending the above studies to the 3D case, in \cite{conetal2016s} we 
considered 3D harmonic oscillator models of the Hamiltonian form
\begin{equation}\label{hamhar3d}
H = {p_1^2\over 2m_1}+{p_2^2\over 2m_2}+{p_3^2\over 2m_3} 
+{1\over 2}(\omega_1^2x_1^2+\omega_2^2x_2^2+\omega_3^2x_3^2) 
\end{equation}
with wavefunctions given by the superposition of three eigenstates
\begin{eqnarray}\label{psi3d3}
\Psi_{n_1,n_2,n_3}(x_1,x_2,x_3,t)&=&e^{-iE_{n_1,n_2,n_3}t/\hbar} \\
&\times&\prod_{k=1}^3\left({m_k\omega_k\over\hbar\pi}\right)
\left({1\over 2^{n_k}n_k!}\right)^{1/2}
\exp\left(-{m_k\omega_kx_k^2\over 2\hbar}\right)
H_{n_k}\left(\sqrt{m_k\omega_k\over\hbar} x_k\right) \nonumber
\end{eqnarray}
with particular combinations of the three integers $n_k$, $k=1,2,3$ and 
$$
E_{n_1,n_2,n_3} = \sum_{k=1}^3 \hbar\omega_k(1/2 +n_k)~~.
$$
The wavefunction takes the form
\begin{equation}\label{wf3d}
\Psi(t) = a\Psi_{p1,p2,p3}(t)+b\Psi_{r1,r2,r3}(t)+c\Psi_{s1,s2,s3}(t)~~. 
\end{equation}
Then we find that for particular combinations of the integers $p_k$, $r_k$ 
and $s_k$ the Bohmian equations of motion admit a conserved quantity, i.e. 
an exact integral of motion. For example, in the case $\omega_1=1$, 
$\omega_2=\sqrt{2}$, $\omega_3=\sqrt{3}$, if we set $(p_1,p_2,p_3)=(1,0,0)$, 
$(r_1,r_2,r_3)=(0,1,0)$ and $(s_1,s_2,s_3)=(0,0,2)$, the Bohmian trajectories 
obey the integral
\begin{equation}\label{cint}
C = x_1^2 + x_2^2 + {1\over 2}x_3^2 -{\sqrt{3}\over 6}\ln|x_3| ~~.
\end{equation}

\begin{figure}
\centering
\includegraphics[scale=0.25]{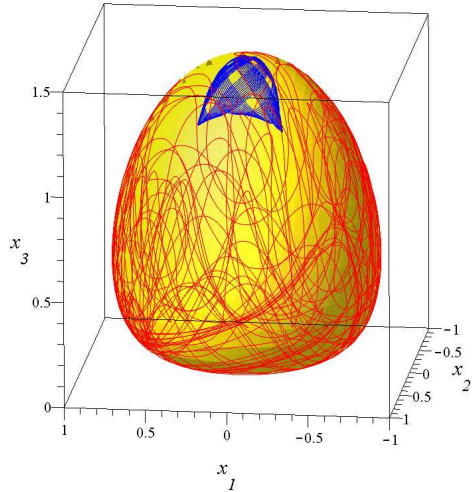}
\caption{Two trajectories in the model (\ref{wf3d}) with $(p_1,p_2,p_3)=(1,0,0)$, 
$(r_1,r_2,r_3)=(0,1,0)$ and $(s_1,s_2,s_3)=(0,0,2)$, which remain confined 
on the surface $C = x_1^2 + x_2^2 + {1\over 2}x_3^2 -{\sqrt{3}\over 6}\ln|x_3|$ 
with $C=2.39255$. The blue trajectory is ordered while the red is chaotic. }
\label{fig:trajint3d}
\end{figure}
Such an integral restricts all the Bohmian trajectories to move on invariant 
surfaces of a given constant value of $C$, as illustrated in Fig.\ref{fig:trajint3d}. 
The trajectories on this surface can be either regular or chaotic. The surface 
is itself intersected transversally by nodal lines, composed of nodal points 
which move in time in each invariant surface. The orthogonal planes defined in 
subsection (2.2) are tangent to the family of surfaces (for different values 
of $C$), and a nodal point - X-point complex is formed in each of these planes. 
Consequently, the Bohmian trajectories which never approach the X-points are 
ordered, while if they approach one or more X-points they are chaotic. 
Nevertheless, even the chaotic trajectories are now fully confined on invariant 
surfaces and they cannot develop any motion in the direction tranversally to 
each trajectory's corresponding invariant surface. 

In \cite{conetal2016s} we explore several cases where exact invariants of motion 
can be constructed by an appropriate combination of the integers $p_k,r_k,s_k$. 
For example, invariants exist for all combinations of the type: 
($p_1 = r_1, r_2 = s_2, s_3 = p_3$), ($r_1 = p_1, s_2 = p_2, s_3 = r_3$), 
($s_1 = r_1, r_2 = p_2, s_3 = p_3$), ($s_1 = r_1, s_2 = p_2, r_3 = p_3$), 
($s_1 = p_1, r_2 = p_2, s_3 = r_3$), ($s_1 = p_1, s_2 = r_2 r_3 = p_3$). 
Further examples are given in \cite{conetal2016s}.

\section{Dynamics of quantum relaxation}

As was already mentioned, the theory of quantum relaxation \cite{val1991} 
constitutes an extension of ordinary quantum mechanics, which, besides 
quantum particles\cite{val1991}\cite{valwes2005}\cite{eftcon2006}\cite{ben2010}
\cite{towetal2012}, can be implemented also to quantum fields \cite{colstru2010}
\cite{col2012}. Valentini's H-function
\begin{equation}\label{hf}
H = \int dq \overline{\rho}\ln(\overline{\rho}/\overline{|\psi|^2})
\end{equation}
is defined in terms of the coarse-grained average particle density 
$\overline{\rho}$, which can be chosen initially to differ from the 
the average value of the square-modulus of the wavefunction 
$\overline{|\psi|^2}$ under the same coarse-graining. The simplest 
possible coarse-graining is provided, e.g., by a cartesian grid dividing 
the configuration space in square cells. A necessary condition is that 
the wavefunction presents initially no `micro-fine structure' (see 
\cite{val1991} for precise definitions). Then, it is shown that $H$ 
satisfies the inequality $H(t)-H(0)\leq 0$ at all times $t>0$. As stressed 
in \cite{conetal2012}, this inequality does {\it not} imply a monotonic 
decrease of $H(t)$ over time. Furthermore, even when $H(t)$ decays to  
small values, this does not necessarily imply a cell-by-cell proximity 
of $\rho$ to $|\psi|^2$, since the integrand in (\ref{hf}) can be both
positive or negative. 

For these reasons, in numerical simulations it is preferable to check the 
approach in time of $\rho$ to $|\psi|^2$ directly by measuring the $L_1$ 
norm of the difference between the two quantities in a grid. To this 
end, in our simulations we first define a smoothed coarse-grained density
\cite{eftcon2006}:
\begin{equation}\label{prosm}
P_s(x, y, t)=\sum_{i=1}^{N}
A\exp\left[{(x-x_i(t))^2 + (y-y_i(t))^2\over 2\sigma^2}\right]
\end{equation}
where $\sigma$ is a Gaussian smoothing length set to a value about equal 
to the cell size, and $A$ is a normalization constant. Then, we define 
the density difference measure $D(t)$
\begin{equation}\label{dendif}
D(t)=\sum_{k=1}^N\sum_{l=1}^N
\left| P_s(x_k,y_l,t) - |\psi(x_k,y_l,t)|^2\right|
\end{equation}
where the sum is over all grid points. The density difference 
$D(t)$ can be compared to Valentini's H-function 
\begin{equation}\label{hsig}
H_s(t)=\sum_{k=1}^N\sum_{l=1}^N
P_s(x_k,y_l,t)\log\left(P_s(x_k,y_l,t)/|\psi(x_k,y_l,t)|^2\right)
\end{equation}
by remarking \cite{conetal2012} that 
$$
P_s\log(P_s/|\psi|^2) = P_s-|\psi|^2 + O[P_s(P_s/|\psi^2| -1)^2]~
$$
for all cells satisfying $P_s>|\psi|^2$, which are the ones mainly 
contributing to the sum (\ref{hsig}). Thus, for small fluctuations 
of $P_s$ vs. $|\psi|^2$ one has $D(t)\approx H(t)$. At any rate, 
$D(t)$ directly measures the deviation of the particle density from 
the square of the wavefunction, thus it is the most convenient measure 
for tests of quantum relaxation. 

\begin{figure}
\centering
\includegraphics[scale=0.45]{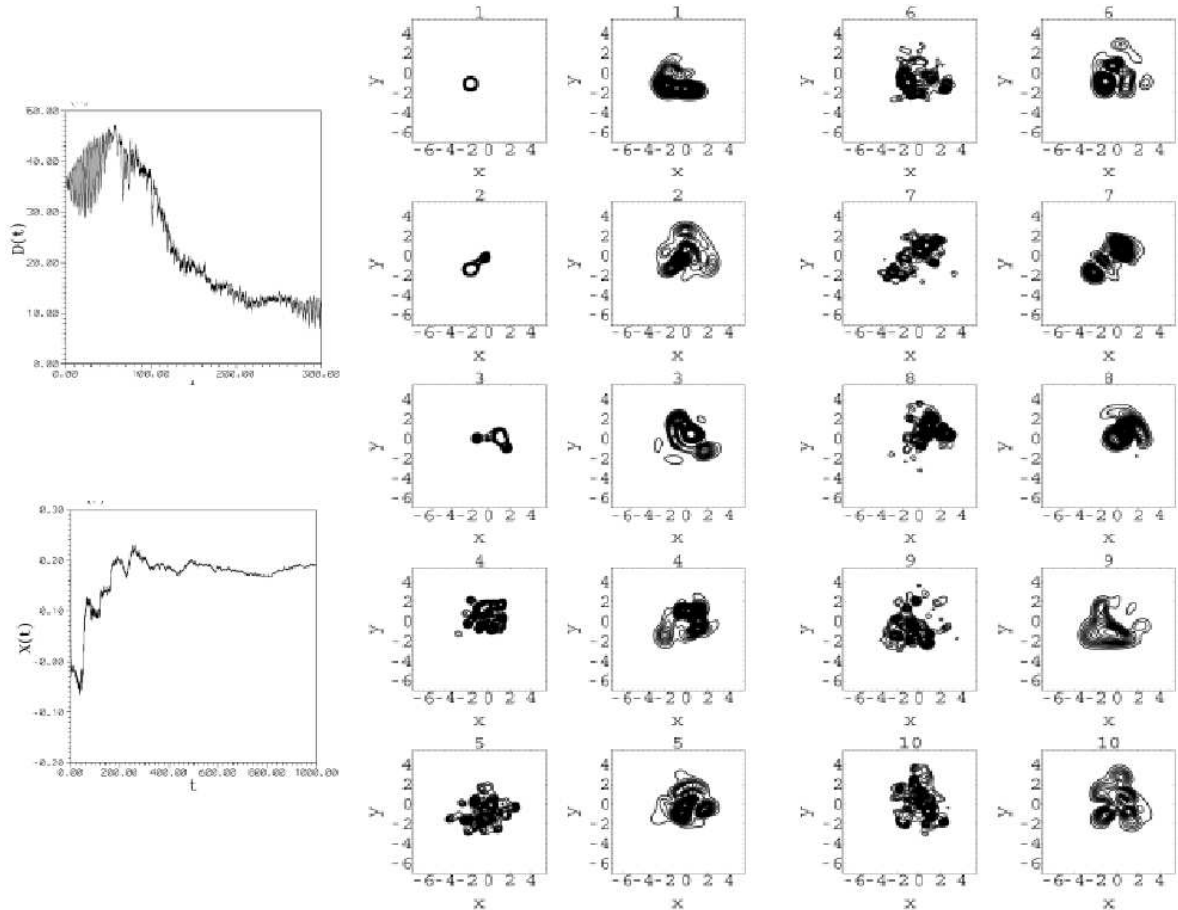}
\caption{Top left: time evolution of the density difference $D(t)$ 
in the model (\ref{hamnl}) with initial wavefunction given by Eq.(\ref{wpckt}), 
when the Bohmian trajectories have an initially homogeneous density in a 
central square box instead of the Gaussian distribution corresponding to 
$|\psi|^2$ (see \cite{eftcon2006}). Bottom left: time evolution of the 
finite-time Luapunov characteristic number $\chi(t)$ for a Bohmian trajectory 
with initial conditions at the center of the initial wavepacket. Right array: 
Contour plots showing the time evolution of the smoothed density $P_s$ 
(first and third column) vs. $|\psi|^2$ (second and fourth column). }
\label{fig:reltrue}
\end{figure}

Figure \ref{fig:reltrue}, compiled from results in \cite{eftcon2006}, 
shows an example of how the quantum relaxation proceeds in time in an 
example of a wavepacket propagating in a Hamiltonian model of harmonic 
oscillators with non-linear coupling:
\begin{equation}\label{hamnl}
H={1\over 2}(p_x^2 + p_y^2 + x^2 + y^2) +\lambda x(y^2-{1\over 3}x^2)~~.
\end{equation}
We follow in time the Schr\"{o}dinger evolution of the initial wavepacket 
\begin{equation}\label{wpckt}
\psi(x,y,t) = \psi_0\exp(-{1\over 2}[(x-x_0)^2 + (y-y_0)^2 
+ i(p_{x0}x +p_{y0}y)]
\end{equation}
which, for $\lambda=0$ represents a so-called (Glauber) `coherent state' of 
the 2D harmonic oscillator. However, due to the nonlinear coupling, the wavepacket's 
spatial coherence is gradually lost. The second and fourth column in the right 
array of panels of Fig.\ref{fig:reltrue} show contour plots of the wavefunction 
in the plane $(x,y)$ at different time snapshots. We observe that the wavefunction 
exhibits an overall higher degree of complexity, with fluctuations taking place 
in smaller and smaller scales as the time increases. 

The first and third column, now, show the contour plots of the particle density 
$P_s$ when this is initially set to be very different from the square of the 
wavefunction (\ref{wpckt}), namely we choose an initial uniform distribution 
in the central cells of the grid. We observe, however, that the density $P_s$ 
progressively evolves in space approaching closer and closer to $|\psi|^2$. 
The top left panel measures the time evolution of the density difference 
$D(t)$ which is observed to decrease in time following an initial time interval 
in which it remains at relatively high levels. The lower left panel shows the 
time evolution of the Lyapunov number for one Bohmian trajectory with initial 
conditions at the center of the initial wavepacket. The Lyapunov number stabilizes 
in time at a positive value, i.e., this trajectory is chaotic. In fact, 
we find that the whole wavepacket is composed by a swarm of such chaotic 
trajectories. 

\begin{figure}
\centering
\includegraphics[scale=0.4]{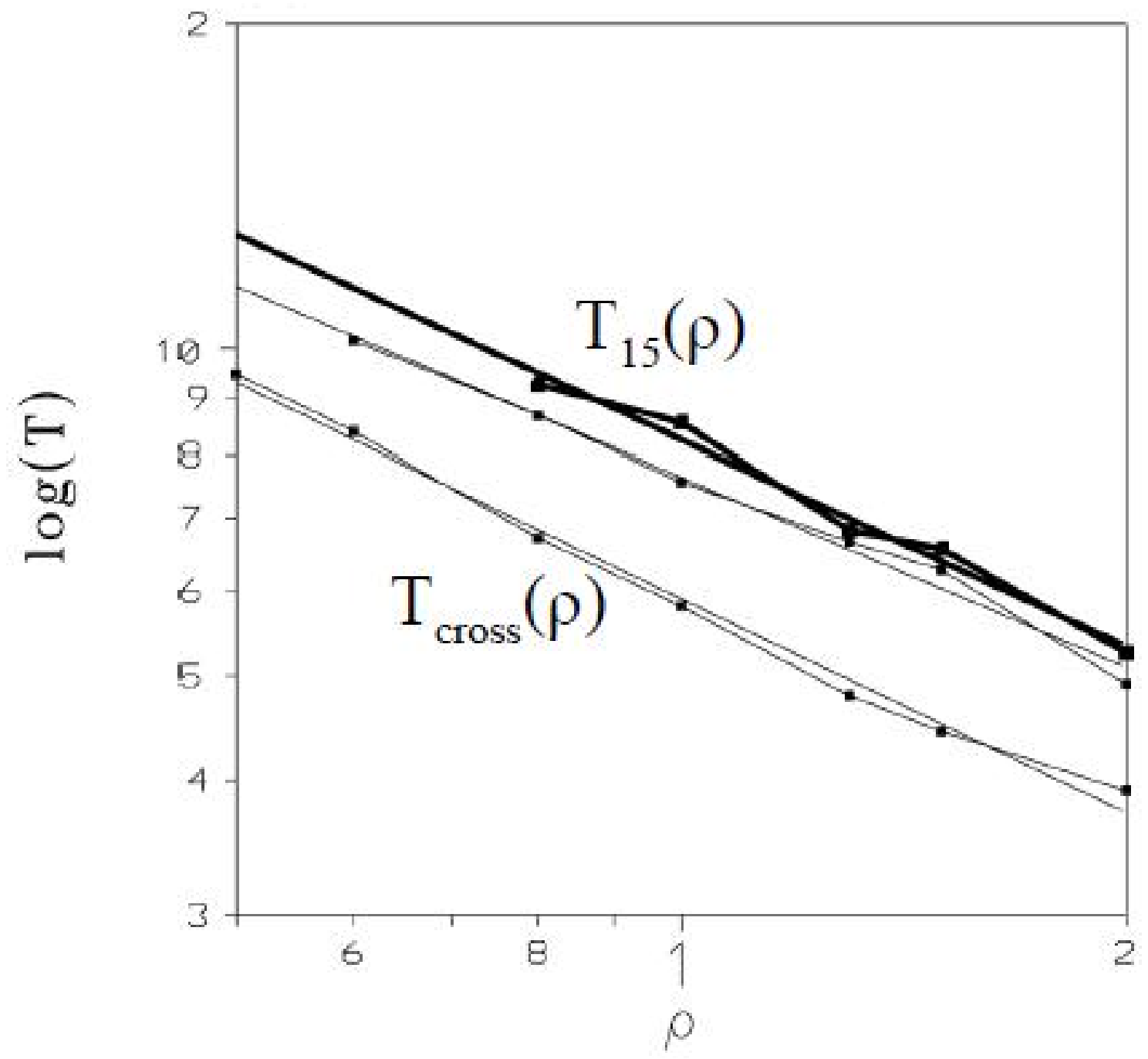}
\caption{The time $T$ it takes for $P_s$ to converge to $|\psi|^2$ in the 
same example as in Fig.\ref{fig:reltrue}, measured by two independent 
indicators $T_{cross}$ and $T_{15}$ (see \cite{eftcon2006} for details), 
as function of the parameter $\rho$ which measures the average energy of 
a wavepacket.}
\label{fig:nekho2}
\end{figure}
In \cite{eftcon2006} we quantified the rate at which the quantum relaxation 
advances in the case of wavepackets as above, by measuring the time it takes 
for $D(t)$ to significantly depart from the initial `plateau' which measures 
the initial deviation of $P_s$ from $|\psi|^2$. As shown in Fig.\ref{fig:nekho2}, 
the time $T$ (defined by various estimates, see \cite{eftcon2006}) follows 
a Nekhoroshev-type result \cite{nekh1977}, i.e., it is exponentially long in 
the inverse of the parameter $\rho=(x_0^2 + y_0^2 + p_{x0}^2 + p_{y0}^2)^{1/2}$, 
where, according to Eq.(\ref{wpckt}) the constants $x_0,y_0,p_{x0},p_{y0}$ define 
the center of the wavepacket in position and momentum space at $t=0$. The 
quantity $\rho$ corresponds, in the classical limit, to a mean energy of the 
wavepacket $\epsilon\sim \rho^2$. We find, by fitting, that 
$T\sim\exp(1/\rho^{0.6})$. The relation between Nekhoroshev stability and 
`quantum normal forms' is discussed in \cite{eft2015}. 

\begin{figure}
\centering
\includegraphics[scale=0.35]{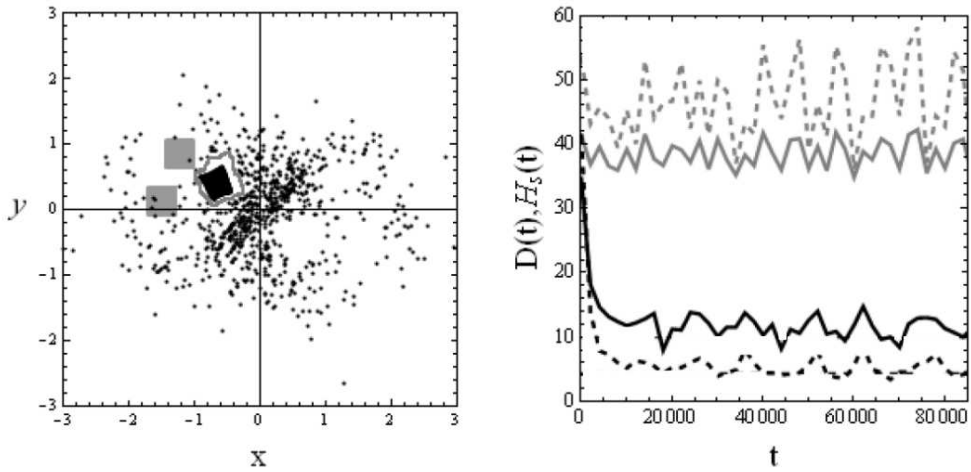}
\caption{Left panel: the gray square boxes show the domains of initial 
conditions for two different sets of trajectories, with data as in figure 10 
of \cite{conetal2012}. The black points give the images of these two boxes at 
a later time $t=10^4$, through the time evolution of the corresponding Bohmian 
trajectories. The upper gray square yields regular orbits which exhibit no 
dispersion in configuration space (black square), while the lower gray box 
corresponds to chaotic orbits which fill a large domain in configuration space 
at later times. Right panel: time evolution of the quantities $D(t)$ (solid), 
and $H_s(t)$ (dashed) for the regular trajectories (top curves, gray) and 
for the chaotic trajectories (lower curves, black).}
\label{fig:rel002011}
\end{figure}
The relevance of the degree of chaos to the rate of quantum relaxation 
can be seen in systems with mixed (i.e. co-existing regular and chaotic) 
dynamics. For example, in \cite{conetal2012} we considered the wavefunction model:
\begin{equation}\label{psi002011}
\psi(x,y,t))=e^{-{x^2+cy^2\over 2}}e^{-{1+c\over 2}it}
\left[1+a(x^2-1)e^{-2it}+bc^{1/2}xye^{-it}\right]
\end{equation}
for $a=1.23$, $b=1.15$, $c=\sqrt{2}/2$. Figure \ref{fig:rel002011} shows results 
for the quantum relaxation with two sets of trajectories, taken with initial 
conditions inside a $0.4\times 0.4$ box centered at i) $x_0=-1.5$, $y_0=0.1275$
and ii) $x_0=-1.23$, $y_0=0.84$. The central values (i) and (ii) were chosen 
so as to correspond to a regular and a chaotic trajectory respectively. 
We observe that the regular trajectories do not exhibit the mixing behavior 
in configuration space, in fact they simply move as a coherent square of nearly 
equal area at all times $t$. Furthermore, the boundary of the area occupied by 
these trajectories cannot be penetrated by chaotic trajectories, despite the 
fact that the chaotic trajectories occupy a much larger area. These differences 
result in a different rate of quantum relaxation for the two ensembles of 
trajectories, as shown in Fig.\ref{fig:rel002011}b. In fact, in the case of 
regular trajectories, both quantities $D(t)$ and $H(t)$ remain in time bounded 
away from zero, while for the chaotic ensemble they both tend to lower asymptotic 
values $D\simeq 12.3$, and $H_s\simeq 6.4$. As argued in \cite{conetal2012}, 
these values, albeit low, are also bounded away from zero, a fact related 
to the non-penetration of the chaotic trajectories within the domains occupied 
by regular trajectories. In other words, in systems in which regular and 
chaotic trajectories co-exist, we find that the quantum relaxation is 
never complete. More examples of various types leading to the same 
conclusion were presented in \cite{conetal2012}, while similar conclusions 
were drawn in an independent subsequent study \cite{eitetal2014}.

\section{Conclusions}
In the present paper we review the topic of the emergence of chaos 
in the de Broglie - Bohm picture of quantum mechanics, and we discuss the role 
that chaos plays in the dynamical origin of the phenomenon of `quantum relaxation'. 
Our main conclusions are:

1) We give a summary of our research on a generic mechanism producing chaotic 
Bohmian trajectories, based on the notion of `nodal point - X-point complex'. 
In particular, we discuss the general structure of the quantum flow near quantum 
vortices, and we explain how the Bohmian trajectories are scattered by the `X-point' 
which exists near every moving nodal point of the wavefunction. This mechanism 
results in a sensitive dependence of the trajectories on the initial conditions, 
leading to positive values of the Lyapunov characteristic exponent. 

2) Scaling laws relating the value of the trajectories' local Lyapunov exponents 
to the size and speed of the nodal point - X-point complexes can be derived 
by considering generic expansions of the wavefunction, and of its associated 
Bohmian equations of motion, around each moving nodal point. 

3) The notion of nodal point - X-point complexes can be extended from 2D to 3D 
systems. In the 3D case, the nodal points belong to continuous families, which 
form `nodal lines' in the 3D configuration space. A 3D cylindrical structure of 
nodal point - X-point complexes is formed around every nodal line. The trajectories 
become chaotic whenever they have close encounters with the X-points along one 
or more such structures. 

4) Besides chaos, we discussed several examples of ordered Bohmian 
trajectories. These are trajectories which avoid coming close to  
any nodal point - X-point complex. Ordered trajectories can be represented 
by series solutions of the Bohmian equations of motion in powers of a suitably 
defined small parameter. In the 3D case we also identify examples of partially 
integrable systems, in which the Bohmian trajectories obey some explicit integral 
of motion. We demonstrate that partial integrability necessarily restricts all 
the Bohmian trajectories into invariant 2D surfaces, whose definition can 
be given in closed form via the corresponding integrals of motion. 

5) We finally discuss how the relative degree of order or chaos in a quantum system 
affects the phenomenon of quantum relaxation, i.e., the approach in time to Born's 
rule $p=|\psi|^2$ for ensembles of Bohmian trajectories with initial conditions 
assumed to deviate from this rule. In this respect, we emphasized that chaos is a 
necessary condition for a system to dynamically approach the state of `quantum 
equlibrium' (i.e. for $p$ to approach in time the value of $|\psi|^2$ everywhere 
in space). In fact, through the study of chaos we were able to specify both 
examples and counter-examples of the effectiveness of quantum relaxation, 
and to quantify the rate of approach of a system to quantum equilibrium, 
as related to the system's underlying level of chaotic behavior. \\
\\
\noindent{\bf Acknowledgements:} This research was suported in part by the 
Research committee of the Academy of Athens. C.E. gratefully acknowledges the 
hospitality of the organizer of the Marseille workshop Dr. T. Durt. \\
\\


\end{document}